\documentclass[12pt]{amsart}
\usepackage[T1]{fontenc}
\usepackage{hyperref}
\usepackage{pstricks}
\usepackage{xspace}
\usepackage{epigraph}
\usepackage{pstricks,pst-node,pst-tree,multido}
\newcommand{\seqnum}[1]{\href{http://oeis.org/#1}{\underline{#1}}}
\newcommand{\teqnum}[1]{\textrm{\seqnum{#1}}}
\newcommand{\U}{\ensuremath{\mathop{\color{blue}{\uparrow}}}\xspace} 
\newcommand{\D}{\ensuremath{\mathop{\color{blue}{\rightarrow}}}\xspace} 

\title{1700 Forests}
\author{Nachum Dershowitz}
\date{\today}                                           
\begin{document}
\maketitle
\thispagestyle{empty}
\epigraph{\it Arvoles lloran por lluvia\\
y montañas por aire \\
Ansí lloran mis ojos \\
por ti querido amante}{Ladino folk song}

\begin{abstract}
Since ordered trees and Dyck paths  are equinumerous, so are
ordered forests and grand-Dyck paths that start with an upwards step.
\end{abstract}

\section{Introduction}

\begin{figure}[t]
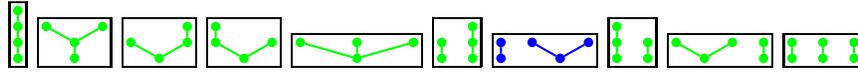

\centering
\psset{treemode=U,radius=2pt,levelsep=6pt,linecolor=green}
\fbox{\pstree{\Tdot}{\pstree{\Tdot}{\pstree{\Tdot}{\Tdot}}}}
\fbox{\pstree{\Tdot}{\pstree{\Tdot}{\Tdot\Tdot}}}
\fbox{\pstree{\Tdot}{\Tdot\pstree{\Tdot}{\Tdot}}}
\fbox{\pstree{\Tdot}{\pstree{\Tdot}{\Tdot}\Tdot}}
\fbox{\pstree{\Tdot}{\Tdot\Tdot\Tdot}}
\fbox{\pstree{\Tdot}{\Tdot}\quad\pstree{\Tdot}{\pstree{\Tdot}{\Tdot}}}
\fbox{\psset{linecolor=blue}\pstree{\Tdot}{\Tdot}\quad\pstree{\Tdot}{\Tdot\Tdot}}
\fbox{\pstree{\Tdot}{\pstree{\Tdot}{\Tdot}}\quad\pstree{\Tdot}{\Tdot}}
\fbox{\pstree{\Tdot}{\Tdot\Tdot}\quad\pstree{\Tdot}{\Tdot}}
\fbox{\pstree{\Tdot}{\Tdot}\quad\pstree{\Tdot}{\Tdot}\quad\pstree{\Tdot}{\Tdot}}
\caption{Ten triple-edge forests ($n=3$).}\label{fig:forests}
\end{figure}

We are interested in the number of  ordered forests (that is, sequences of non-trivial ordered rooted trees) with a total of $n$ edges.
Every tree in the forest must have at least one edge
(in that sense they are non-trivial), or else there would be infinitely many forests for every $n$.
For example, there are $\binom{5}{3} =10$ such forests with $n=3$ edges, as depicted in Figure~\ref{fig:forests}.

It turns out---easily enough---that
these forests are counted by
\begin{align}\label{eq}
F_n &= \frac{1}{2}\binom{2n}{n} = \binom{2n-1}{n} 
\end{align}
This enumeration is sequence \seqnum{A001700} in Neil Sloane's \textit{On-Line Encyclopedia of Integer Sequences (OEIS)}:%
\footnote{\url{http://oeis.org}. Previously in print form: Neil J. A. Sloane, \textit{A Handbook of Integer Sequences}, Academic Press, NY, 1973.}
\[
\begin{array}{r|cccccccccc}
n & 0 & 1 & 2 & 3 & 4 & 5 & 6 & 7 & 8 & \cdots\\\hline
F_{n+1} & 1 & 3 & 10 & 35 & 126 & 462 & 1716 & 6435 & 24310 &\cdots
\end{array}
\]
In other words, sequence $\teqnum{A001700}(n)=F_{n+1}$ also counts the number of ordered forests with $n+1$ edges (and no trivial trees).

Compare this forest enumeration with the Catalan numbers, $C(n)=\frac{1}{n+1}\binom{2n}{n}$, which count (among many combinatorial objects)
ordered forests with $n$ \textit{nodes}, allowing for trivial (leaf-only, edgeless) trees.
They form sequence \seqnum{A000108} in the \textit{OEIS}.

Perhaps it is
because  parameterizing by the number of edges  is less common than
the use of a node parameter that this enumeration of ordered forests has not appeared  in the literature until now.

\section{Paths and Forests}

The justification for enumeration (\ref{eq}) follows from the standard correspondence between trees and lattice paths.
A \emph{Dyck path}
is a (monotonic, ``staircase'') lattice path (consisting of a mix of \U and \D steps)
beginning and ending on the diagonal and never venturing below;
a \emph{grand-Dyck} path may go both above and below the diagonal but must end on it.%
\footnote{Grand-Dyck paths are classified as ``bridges'' in 
Cyril Banderier and Philippe Flajolet,
``Basic analytic combinatorics of directed lattice paths'', \textit{Theoretical Computer Science} \textbf{281} (2002): 37--80.}

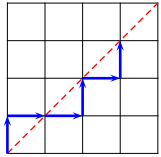
\begin{figure}[t]
\centering
\psscalebox{0.5}{
\begin{pspicture}[showgrid=false,unit =1cm](0,0)(4,4)
\psgrid[griddots=0,gridwidth=0.1pt,subgriddiv=1,gridlabels=0]
\psline[linewidth=1pt,linestyle=dashed,linecolor=red](0,0)(4,4)
\psline[linewidth=2pt,linecolor=blue,arrows=->](0,0)(0,1)
\psline[linewidth=2pt,linecolor=blue,arrows=->](0,1)(1,1)
\psline[linewidth=2pt,linecolor=blue,arrows=->](1,1)(2,1)
\psline[linewidth=2pt,linecolor=blue,arrows=->](2,1)(2,2)
\psline[linewidth=2pt,linecolor=blue,arrows=->](2,2)(3,2)
\psline[linewidth=2pt,linecolor=blue,arrows=->](3,2)(3,3)
\end{pspicture}}
\caption{The grand-Dyck path corresponding to the 7th (blue) forest in Figure~\ref{fig:forests}.}\label{dyck}
\end{figure}

Ordered (rooted plane) trees with $n$ edges are well-known to be in bijection with Dyck paths of length $2n$.%
\footnote{David A. Klarner, ``Correspondence between plane trees and binary sequences'', \textit{Journal of Combinatorial Theory} \textbf{9} (1970) 401--411.}
So a forest, which is a sequence of trees, corresponds to a sequence of Dyck paths.
Every 
grand-Dyck can be interpreted as a sequence of Dyck paths, one per (non-trivial) tree,
delineated by the points at which the path crosses the diagonal.
See Figure~\ref{dyck} (left).
There are $2n \choose n$ such paths (since they must have $n$ \U steps and $n$  \D steps).
But a path and its mirror image (reflected about the diagonal) correspond to the same forest.
The equation follows.

\section{Height Restrictions}

The correspondence between forests and grand-Dyck paths applies equally to trees of restricted height
and paths within a band, the latter analyzed by Mohanty.%
\footnote{Sri Gopal Mohanty, \textit{Lattice Path Counting and Applications}, volume 37 of \textit{Probability and Mathematical Statistics}, Academic Press, New York, 1979, pp.\@ 6--7.}
It follows that the number of $n$-edge forests whose trees are all of height at most $h$ is
\begin{align}
F_n^h &= \frac12 \sum_{k\in\mathbb{Z}} \left[\binom{2n}{n+2k(h+1)}-\binom{2n}{n+(2k+1)(h+1)}\right] 
\end{align}

For example, there are
\begin{align*}
F_3^1 &= \frac12 \sum_k \left[\binom{6}{3+4k}-\binom{6}{5+4k}\right]\\
& =
\frac12 \left[\binom{6}{-1}-\binom{6}{1}+\binom{6}{3}-\binom{6}{5}\right]\\
&=\frac12[0-6+20-6]=4
\end{align*}
forests in Figure~\ref{fig:forests} with trees of height 1.

\end{document}